%Paper: hep-th/9408028
%From: YVONNE@urhep.pas.rochester.edu
%Date: Thu, 04 Aug 1994 16:02:17 -0500 (EST)

\baselineskip=12pt
\magnification=1200
\tolerance=100000
\overfullrule=0pt

\rightline{UR-1366\ \ \ \ \ \ \ }

\rightline{ER-40685-816}

\baselineskip=18pt

\smallskip

\centerline{\bf A SUPERSPACE FORMULATION OF THE BV ACTION}
\vskip 1 true cm
\centerline{Nelson R. F. Braga }
\bigskip
\centerline{and }
\bigskip
\centerline{Ashok Das$^\ast$}

\vskip .5 true cm
\centerline{Instituto de F\'isica}
\centerline{Universidade Federal do Rio de Janeiro}
\centerline{Rio de Janeiro  21945  Caixa Postal 68.528}
\centerline{Brasil}
\vskip 1.5 true cm
\centerline{\bf Abstract}
\vskip 0.5 true cm

 We show that the BV (Batalin Vilkovisky) action, formulated
with an extended
BRST symmetry (including the shift symmetry), is also invariant under
an extended anti-BRST transformation (where the antifields are the
parameters of the transformation), when the gauge fixing Lagrangian
is both BRST and anti-BRST invariant. We show that for a general
gauge fixing Lagrangian, the BV action can be written in a manifestly
extended BRST invariant manner in a superspace with one Grassmann coordinate
whereas it can be expressed in a manifestly extended BRST and
anti-BRST invariant manner in a superspace with two Grassmann
coordinates when the gauge fixing Lagrangian is invariant under both
BRST and anti-BRST transformations.

\vskip 1 true cm
\noindent $\ast$ Permanent address: Department of Physics and Astronomy,
University of Rochester, Rochester, NY 14627, USA.
\vskip 0.5cm

\vfill\eject

\noindent{\bf I. \underbar{Introduction}}

\medskip

The Batalin-Vilkovisky (BV) formalism [1,2] has proved to be a powerful
method of quantization for various gauge theories, as well as
supergravity theories and topological field theories in the
Lagrangian framework. It encompasses the Faddeev Popov quantization
and uses the BRST symmetry, discovered in the context of gauge
theories [3], to build on it. It is known that, for standard gauge
fixing, gauge theories possess an anti-BRST invariance [4] in
addition to the usual BRST symmetry. In contrast, however, the
standard discussion of the BV formalism has so far been predominantly
in the BRST context [5]. Furthermore, while the BRST and the anti-BRST
symmetry of gauge theories can be given a geometrical meaning and
have led to a superspace formulation of such theories [6,7],
a superspace
description of the BV action does not exist so far. The main
difficulty behind this lies in the fact that the antifields,
introduced in the BV formalism, are identified with the functional
derivative of a gauge fixing fermion with respect to the associated
field. Thus, a priori, it would appear that a superspace description
cannot even be contemplated before choosing a gauge.

Recently, however, it has been shown [8] how an extended BRST invariant
formulation (including the shift symmetry) of the BV action,
naturally leads to the proper identification of the antifields
through equations of motion of auxiliary field variables. This,
therefore raises the possibility of a superspace formulation of the
BV action which we discuss in this paper. In sec. II we review the
extended BRST invariant formulation of the BV action. In sec. III we
show, when the gauge fixing Lagrangian is both BRST and anti-BRST
invariant, that this action possesses an extended anti-BRST
invariance, where the antifields define the transformations.  In sec.
IV we show how, for a general gauge fixing, the BV action can be
written in superspace with a manifest extended BRST invariance.  In
sec. V we use both the extended BRST and anti-BRST invariances to
formulate the BV action in superspace with two Grassmann coordinates
and present our conclusions in sec. VI . For clarity of ideas, our
entire discussion will be in the context of a non Abelian Yang Mills
theory with covariant gauge fixing. The generalization to other
systems can be carried out in a straight forward manner.
\vskip 1cm
\noindent{\bf II. \underbar{Extended BRST Invariant Formulation}}

\medskip

Let us consider a non Abelian Yang Mills theory described by:

$$ {\cal L} = {\cal L} (A_\mu, c, \overline c, F ) =
{\cal L}_o(A_\mu ) + {\cal L}_{gf} (A_\mu, c, \overline c, F )\eqno(2.1)$$

\noindent where ${\cal L}_o(A_\mu )$ describes the gauge invariant classical
Lagrangian density, while $ {\cal L}_{gf}$ represents the Lagrangian density
associated with gauge fixing and ghosts. We assume that all the
fields -- the gauge field, ghost, antighost, and the auxiliary field --
are matrices belonging to the adjoint representation of the gauge
group which we can choose to be SU(n). Thus, for example, for a
standard covariant gauge fixing, we have

$$\eqalign{&{\cal L}_o(A_\mu ) = -{1\over 4}\  {\rm Tr}\  F_{\mu\nu} F^{\mu\nu}
\cr
\noalign{\vskip 4pt}%
&{\cal L}_{gf} = \ {\rm Tr}\  ( \partial_\mu F A^\mu + \partial_\mu\overline c
D^\mu c )\cr}\eqno(2.2)$$

\noindent where

$$\eqalign{D_\mu c &= \partial_\mu c  + [ A_\mu , c ]\cr
 F_{\mu\nu} &= \partial_\mu A_\nu - \partial_\nu A_\mu
+[A_\mu , A_\nu ] \cr}\eqno(2.3)$$

\noindent and Tr stands for the trace over the SU(n) indices.

As we know, the Lagrangian density in (2.1) is invariant under the
BRST transformations

$$\eqalign{&\delta A_\mu = \omega D_\mu c \cr
&\delta c = - {\omega\over 2}\  [c,c]_+\cr
&\delta \overline c = -\omega F \cr
&\delta F = 0  \cr}\eqno(2.4)$$

\noindent where $\omega$ is a constant Grassmann parameter. In the BV
formalism ${\cal L}_{gf}$ is represented as a BRST variation of a
general fermion which results in the BRST invariance of the theory.
It can be easily checked that the choice of gauge fixing in (2.2)
can also be represented as (without the BRST parameter)

$$ {\cal L}_{gf} = \delta \psi \eqno(2.5)$$

\noindent where the gauge fixing fermion has the form

$$\psi = - {\rm Tr}\  \partial_\mu \overline c A^\mu \eqno(2.6)$$

It is straight forward to check that the Lagrangian in Eq. (2.1)
is also invariant under the anti-BRST tranformations [4]

$$\eqalign{&\overline \delta A_\mu = \overline \omega D_\mu
\overline c \cr
&\overline \delta c =  \overline\omega ( F - [c,\overline c]_+ )
\cr
&\overline\delta\  \overline c = -
{\overline\omega\over 2}\  [\overline c,\overline c]_+\cr
&\overline\delta F = \overline \omega [ F, \overline c ]  \cr}\eqno(2.7)$$

\noindent where $\overline \omega$ is the constant Grassmann parameter of
the transformation.
We note that the gauge fixing fermion in Eq. (2.6) can be written as
(without the parameter)

$$ \psi = -\, {1\over 2}\  \overline \delta \  {\rm Tr}\
A_\mu A^\mu \eqno(2.8)$$

\noindent This, of course, implies that the Lagrangian is both BRST and
anti-BRST invariant since the transformations in Eqs. (2.4) and (2.7)
are nilpotent.  However, we also note that not all gauge fixing
fermions which lead to BRST and anti-BRST invariance can be written
as the anti-BRST variation of a boson. Furthermore, not all gauge
fixing Lagrangians will be invariant under both BRST and anti-BRST
transformations [9].

The extended BRST invariant formulation of the BV action is obtained
by considering the Lagrangian density [8]

$$ \hat {\cal L} = {\cal L} (A_\mu - \tilde A_\mu, c - \tilde c,
 \overline c -  \tilde{\overline c}, F - \tilde F )
\eqno(2.9)$$

\noindent which coincides with Eq. (2.1) when all the tilde fields vanish.
This Lagrangian density is, of course, invariant under the BRST
transformation of (2.3) with respect to the fields $\phi - \tilde
\phi $ where $ \phi = \{ A_\mu , c , \overline c , F \} $.
But in addition, it is also invariant under the local shift symmetry

$$\eqalign{ \delta \phi (x) &= \alpha (x)\cr
\delta \tilde \phi (x) &= \alpha (x)\cr}\eqno(2.10)$$

\noindent which needs to be gauge fixed and, in turn, leads to an
additional BRST symmetry. Together, the BRST symmetries are commonly
referred to as an extended BRST symmetry with the transformations:

$$\eqalign{&\delta A_\mu = \omega \psi_\mu
\;\;\; , \;\;\;
\delta \tilde A_\mu = \omega (\psi_\mu -  D_\mu^{(A-\tilde A)}
(c - \tilde c)) \cr
&\delta c = \omega \epsilon\;\;\; , \;\;\;
\delta \tilde c = \omega \bigg( \epsilon + {1\over 2}\
[c- \tilde c,c - \tilde c]_+ \bigg) \cr
&\delta \overline c = \omega \overline \epsilon \;\;\;\; , \;\;\;\;
\delta \tilde {\overline c} =  \omega ( \overline\epsilon +
(F - \tilde F) ) \cr
&\delta F = \omega \epsilon_F \;\;\; , \;\;\;
\delta \tilde F = \omega\epsilon_F  \cr
&\delta \psi_\mu = 0\cr
&\delta \epsilon = 0 \cr
&\delta \overline \epsilon = 0\cr
&\delta \epsilon_F = 0\cr
&\delta A^*_\mu = - \omega B_\mu \cr
&\delta B_\mu = 0 \cr
&\delta c^* = - \omega B \cr
&\delta B = 0\cr
&\delta \overline c^* = - \omega \overline B\cr
&\delta \overline B = 0\cr
&\delta F^* = - \omega B_F \cr
&\delta B_F = 0\cr} \eqno(2.11)$$

\noindent Here $\psi_\mu$, $\epsilon$, $\overline \epsilon$ and
$\epsilon_F$ are the ghost fields associated with the shift
symmetries for $ A_\mu , c , \overline c $ and $F$ respectively,
$ A_\mu^* , c^* , \overline c^* $ and $F^*$ are the respective antighosts
and $B_\mu , B , \overline B $ and $B_F$ represent the corresponding
auxiliary fields. We note that there is a certain amount of
arbitrariness in defining the transformations in Eq. (2.11), but this
is the conventional choice.

If we gauge fix the shift symmetry such that all the tilde fields
vanish, then, of course, we will recover our original theory. This
can be achieved by choosing a gauge fixing Lagrangian of the form:

$$\eqalign{\tilde {\cal L}_{gf} =\ {\rm Tr}\
 \bigg[ &-B_\mu\tilde A^\mu - A_\mu^*
(\psi^\mu - D^{\mu (A-\tilde A)} (c-\tilde c) )\cr
\noalign{\vskip 4pt}%
 &- \overline B \tilde
c + \overline c^* \bigg( \epsilon + {1\over 2}\
 [c-\tilde c,c - \tilde c]_+ \bigg)
\cr
\noalign{\vskip 4pt}%
&+ B \tilde {\overline c} - c^* (\overline \epsilon + (F- \tilde F  ))
+ B_F \tilde F + F^* \epsilon_F \bigg] \cr}\eqno(2.12)$$

It is clear that integrating out the auxiliary fields  $B_\mu , B ,
\overline B $ and $B_F$ would set all the tilde fields equal to zero
and it is also straight forward to check that the gauge fixing Lagrangian,
in Eq. (2.12), for the shift symmetry is invariant under the extended
BRST transformations of Eq. (2.11). In addition, of course, we have
the gauge fixing Lagrangian for the original gauge symmetry (see Eqs.
(2.5) and (2.6)).
Being the extended variation of a fermion, this is also invariant
under the transformations in (2.11) which are nillpotent.  If we
choose the gauge fixing fermion to depend on the original fields
only, then a general gauge fixing Lagrangian for the original
symmetry will have the form (we use left derivatives.)

$$\eqalign{{\cal L} &= \delta \psi =
\ {\rm Tr}\  \bigg( \delta A_\mu\  {\delta \psi \over \delta A_\mu} +
\delta c\  {\delta \psi \over \delta c} +
\delta \overline c\  {\delta \psi \over \delta \overline c} +
\delta F\  {\delta \psi \over \delta F} \bigg) \cr
\noalign{\vskip 4pt}%
&=\  {\rm Tr}\  \bigg( - {\delta \psi \over \delta A_\mu}\  \psi^\mu +
{\delta \psi \over \delta c}\ \epsilon +
{\delta \psi \over \delta \overline c}\ \overline \epsilon -
{\delta \psi \over \delta F}\ \epsilon_F \bigg)
 \cr}\eqno(2.13)$$

\noindent Thus, after integrating out the auxiliary fields which set
the tilde fields to zero, we have

$$\eqalign{{\cal L} &= {\cal L}_o(A_\mu )
+\tilde {\cal L}_{gf} + {\cal L}_{gf}
= -{1\over 4}\ {\rm  Tr}\  F_{\mu\nu} F^{\mu\nu}\cr
&+ \ {\rm Tr}\  \bigg( A^*_\mu D^\mu c  + {1\over 2}\  \overline c^*
[c,c]_+
 - c^* F - \bigg( A^*_\mu +
{\delta \psi \over \delta A_\mu}\bigg) \psi^\mu  \cr
&+ \bigg(\overline c^* + {\delta \psi \over \delta c} \bigg)\epsilon -
\bigg( c^* - {\delta \psi \over \delta \overline c} \bigg) \overline \epsilon +
\bigg( F^* -
 {\delta \psi \over \delta F} \bigg) \epsilon_F \bigg)\cr}\eqno(2.14)$$

If we now integrate out the ghosts associated with the shift symmetry,
it leads to the identification

$$\eqalign{  A^*_\mu &=  - {\delta \psi \over \delta A_\mu} \cr
\noalign{\vskip 4pt}%
\overline c^* &= - {\delta \psi \over \delta c}\cr
\noalign{\vskip 4pt}%
c^* &=  {\delta \psi \over \delta \overline c} \cr
\noalign{\vskip 4pt}%
F^* &= {\delta \psi \over \delta F} \cr}\eqno(2.15)$$

For the covariant gauge fixing of Eq. (2.6), this yelds

$$\eqalign{  A^*_\mu &=   \partial_\mu \overline c \cr
\overline c^* &= 0 \cr
c^* &=  \partial_\mu A^\mu\cr
F^* &= 0 \cr}\eqno(2.16)$$

\noindent With the identification of Eq. (2.15) (or (2.16) in the particular
case), we recover the BV action for the theory [1,2].

\vskip 1cm
\noindent{\bf III. \underbar{Extended Anti BRST Invariance}}

\medskip

Let us next consider a gauge fixing Lagrangian (such as in Eq. (2.2))
which is invariant under both the BRST and anti-BRST transformations
of Eqs. (2.4) and (2.7). In this case, the complete Lagrangian is
also invariant under BRST and anti-BRST transformations.
The parameter of transformation for the anti-BRST
transformations (see Eq. (2.7)) is the antighost field.
While the anti-BRST invariance does not lead to any additional
information beyond what BRST invariance provides, it helps put the
theory in a proper geometrical setting. It is, therefore, interesting
to ask whether such a Lagrangian with an extended BRST invariance is
also symmetric under an extended anti-BRST transformation. In fact,
it is stright forward to check that the transformations

\vfill\eject

$$\eqalign{&\overline \delta A_\mu = \overline\omega (A^*_\mu +
D^{(A-\tilde A)}_\mu (c- \tilde{\overline c} ) )
\;\;\; ,\;\;\;
\overline\delta \tilde A_\mu = \overline\omega A^*_\mu \cr
&\overline\delta c = \overline\omega (c^* + (F-\tilde F) -
[c-\tilde c,\overline c - \tilde{\overline c}]_+ )
\;\;\; , \;\;\;
\overline\delta \tilde c = \overline\omega c^*  \cr
&\overline\delta \overline c = \overline\omega (\overline c^*
 - {1\over 2}\  [\overline c-\tilde{\overline c},\overline c -
\tilde{\overline c}]_+ )
 \;\;\; , \;\;\;
\overline\delta \tilde {\overline c} =  \overline\omega \;
\overline c^*  \cr
&\overline\delta F = \overline\omega ( F^* + [F- \tilde F, \overline c
 - \tilde {\overline c}]_+ )
 \;\;\; , \;\;\;
\overline\delta \tilde F = \overline\omega F^* \cr
&\overline\delta \psi_\mu = \overline\omega \;
(B_\mu + D_\mu^{(A-\tilde A)} (F- \tilde F) - [ D_\mu^{(A-\tilde A)}
(c- \tilde c),
\overline c - \tilde{\overline c}]_+ )
\cr
&\overline\delta \epsilon = \overline\omega
(B - [ F- \tilde F, c- \tilde c] + [\overline c - \tilde{\overline c},
(c -\tilde c)^2 ] )\cr
&\overline\delta \overline \epsilon = \overline\omega
(\overline B - [F-\tilde F, \overline c - \tilde{\overline c} ])\cr
&\overline\delta \epsilon_F = \overline\omega B_F\cr
&\overline\delta A^*_\mu = 0  \cr
&\overline\delta B_\mu = 0 \cr
&\overline\delta c^* = 0 \cr
&\overline\delta B = 0\cr
&\overline\delta \overline c^* =0 \cr
&\overline\delta \; \overline B = 0\cr
&\overline\delta F^* = 0 \cr
&\overline\delta \;  B_F = 0 \cr }\eqno(3.1)$$

\noindent define a symmetry of the complete Lagrangian.  The
classical part of the Lagrangian density ${\cal L}_o (A_\mu -
\tilde A_\mu )$, is, of course, invariant under these transformations
simply because they represent a gauge transformation in these
variables.  It is straight forward to check that $\tilde{\cal
L}_{gf}$ is also invariant under these transformations. For a gauge
fixing Lagrangian of the original symmetry which is both BRST and
anti-BRST invariant, it follows that this must be invariant under the
extended anti-BRST transformation at least on-shell where the
transformations reduce to anti-BRST transformations.  For the
covariant gauge fixing choice of Eq. (2.6), invariance on shell is
obvious since the gauge fixing Lagrangian is the anti-BRST variation
of an operator (see Eq.  (2.8)). We will discuss this in more detail
in sec. V.

We note here that, as in the case of extended BRST transformations,
there is an arbitrariness in the transformations in Eq. (3.1).
However, the present choice of transformations leads to a simple
superspace formulation of the theory as we will see in sec. V.

\vskip 1cm
\noindent{\bf IV. \underbar{Extended BRST Invariant Superspace
Formulation}}

\medskip

Let us consider a superspace labelled by the coordinates $(x^\mu,\theta)$
 [6]. In this space, a super-connection 1-form will have the form [7]

$$ \omega =   \phi_\mu (x,\theta) dx^\mu +
\eta (x,\theta) d\theta \eqno(4.1)$$

\noindent where we assume the component superfields to have the form

$$\eqalign{  \phi_\mu (x,\theta) &= A_\mu(x) +
\theta R_\mu(x) \cr
\eta (x,\theta ) &= c(x) + \theta R(x) \cr}\eqno(4.2)$$

\noindent $A_\mu(x)$ and $c(x)$ are assumed to be the gauge fields
and ghosts associated with a Yang Mills theory and all the components
of the superfields in Eq. (4.2) are assumed to belong to the adjoint
representation of the gauge group SU(n). The curvature (field strength)
associated with the connection in Eq. (4.1) is given by

$$ F = d\omega + {1\over 2}\ [\omega , \omega ] \eqno(4.3)$$

It is known [6,7] that if we require the components of the field strength
to  vanish along the $\theta$ direction, then it determines the
superfields to have the form

$$\eqalign{  \phi_\mu (x,\theta) &= A_\mu(x) +
\theta D_\mu c \cr
\noalign{\vskip 4pt}%
\eta (x,\theta) &= c(x) - {1\over 2}\ \theta [c,c]_+ \cr}\eqno(4.4)$$

\noindent In other words, the BRST transformations of $A_\mu(x)$ and
$c(x)$, in this case result as a consequence of translations of the
coordinate $\theta$.  In this formalism, the antighosts and other
matter fields have to be introduced as additional superfields of the
form (for the antighosts)

$$\overline\eta (x,\theta) = \overline c(x)
-\theta F(x) \eqno(4.5)$$

Let us next consider the theory defined by

$$\tilde {\cal L} = {\cal L} (  \phi_\mu - \tilde \phi_\mu\,,\,
\eta - \tilde \eta \,,\,  \overline\eta -  \tilde {\overline\eta} )
\eqno(4.6)$$

\noindent and note that when the tilde superfields vanish, this
Lagrangian reduces to our original theory. In this case, we can define

$$\eqalign{  \Phi_\mu (x,\theta) &=
\phi_\mu (x,\theta) -
\tilde\phi_\mu (x,\theta)\cr
\Lambda (x,\theta) &=
\eta (x,\theta) -
\tilde\eta (x,\theta)\cr }\eqno(4.7)$$

\noindent and note that if the field strength associated with the 1-form

$$\Omega = \Phi_\mu dx^\mu + \Lambda d\theta  \eqno(4.8)$$

\noindent vanishes along the $\theta$ direction, then we can determine

$$\eqalign{  \Phi_\mu (x,\theta) &=
( A_\mu - \tilde A_\mu )  +
\theta  D_\mu^{(A-\tilde A)}( c -\tilde c) \cr
\noalign{\vskip 4pt}%
\Lambda (x,\theta) &= (c - \tilde c)
 - {1\over 2}\ \theta [c- \tilde c,c- \tilde c]_+
 \cr}\eqno(4.9)$$

\noindent This, however, does not determine the individual
superfields $\phi_\mu ,\  \tilde\phi_\mu ,\  \eta$ and $\tilde \eta$ uniquely
and this is the arbitrariness in the extended BRST transformations
that we discussed earlier.

Consistent with the discussion in sec. II, let us choose

$$\eqalign{  \phi_\mu (x,\theta) &=
 A_\mu  + \theta \psi_\mu \cr
\tilde\phi_\mu (x,\theta) &=
 \tilde A_\mu  +
\theta \big(\psi_\mu -  D_\mu^{(A-\tilde A)}( c -\tilde c)\big)\cr
\eta (x,\theta) &= c +
\theta \epsilon \cr
\tilde\eta (x,\theta) &=  \tilde c
+\theta\  \bigg( \epsilon + {1\over 2}\ [c- \tilde c,c- \tilde c]_+
\bigg)\cr
\overline\eta (x,\theta) &=
\overline c
+ \theta \overline\epsilon\cr
\tilde{\overline\eta} (x,\theta) &=
\tilde{\overline c}
+ \theta (\overline\epsilon + (F - \tilde F ) )\cr }\eqno(4.10)$$

\noindent In addition, let us introduce the superfields

$$\eqalign{  \tilde\phi_\mu^* (x,\theta) &=
 A_\mu^*  -
\theta B_\mu \cr
\tilde\eta^* (x,\theta) &= c^* -
\theta B \cr
\tilde{\overline \eta}^* (x,\theta) &=  \overline c^*
-\theta \overline B \cr
 \tilde f (x,\theta) &=
\tilde F + \theta \epsilon_F \cr
\tilde f^* (x,\theta) &=
F^* -
 \theta B_F\cr  }\eqno(4.11)$$

\noindent It is clear now that with these choices of the superfields, the
extended BRST transformations of Eq. (2.11) arise as translations of
the $\theta$ coordinate.

Let us next note the following relations

$$\eqalign{ {\partial\over \partial\theta}\ {\rm  Tr}\
\tilde\phi_\mu^* \tilde\phi^\mu &=
\ {\rm Tr}\  \Big( -B_\mu\tilde A^\mu - A^*_\mu \big( \psi^\mu -
 D^{\mu (A-\tilde A)}
(c-\tilde c) \big) \Big)\cr
\noalign{\vskip 4pt}%
{\partial\over \partial\theta}\ {\rm  Tr}\  \tilde{\overline
 \eta}^*
\tilde \eta &=
\ {\rm Tr}\  \bigg( -\overline B\tilde c +
\overline c^* \bigg( \epsilon + {1\over 2}\
[c-\tilde c, c-\tilde c]_+ \bigg) \bigg) \cr
\noalign{\vskip 4pt}%
-{\partial\over \partial\theta}\ {\rm  Tr}\  \tilde{\overline \eta}
\tilde \eta^* &=\ {\rm  Tr}\ \Big(  B\tilde{\overline c} -
c^* \big( \overline \epsilon + (F-\tilde F)\big) \Big)\cr
\noalign{\vskip 4pt}%
- {\partial\over \partial\theta}\ {\rm  Tr}\  \tilde f^*
\tilde f &=
\ {\rm Tr}\  \Big( B_F \tilde F + F^* \epsilon_F \Big)
 \cr}\eqno(4.12)$$

\noindent Thus, we see that the gauge fixing Lagrangian of Eq. (2.12)
for the shift symmetry can be written in this superspace as

$$\tilde {\cal L}_{gf} =
{\partial\over \partial\theta}\ {\rm  Tr}\  \Big(
\tilde\phi_\mu^*\tilde\phi^\mu
+ \tilde{\overline \eta}^* \tilde\eta -
 \tilde{\overline \eta} \tilde\eta^*
-\tilde f^* \tilde f
\Big) \eqno(4.13)$$

\noindent Being the $\theta$ component of a superfield, this is
manifestly invariant under the extended BRST transformations of
Eq. (2.11).

The gauge fixing Lagrangian for the original symmetry can also be
written in this space in a straight forward manner. Let $\psi = \psi (A_\mu,
c, \overline c, F)$ denote an arbitrary gauge fixing fermion. We
assume this to depend only on the original fields. Then, we can
define a fermionic superfield as

$$ \Psi = \psi +\theta\delta\psi =
\psi + \theta \bigg(- {\delta \psi \over \delta A_\mu}\ \psi_\mu
 +  {\delta \psi \over \delta c}\ \epsilon +
{\delta \psi \over \delta \overline c}\ \overline\epsilon -
{\delta \psi \over \delta F}\  \epsilon_F \bigg)\eqno(4.14)$$

\noindent For the gauge choice in Eqs. (2.5) and (2.6), then, this
superfield will be of the form

$$ \Psi = -{\rm Tr}\  \partial_\mu \overline c A^\mu + \theta
\ {\rm Tr}\  \big(
\partial_\mu\overline c \psi^\mu + \partial_\mu A^\mu \overline
\epsilon \big) \eqno(4.15)$$

\noindent We see that the gauge fixing Lagrangian for the original
symmetry can be written as

$$ {\cal L}_{gf} = {\partial\Psi \over \partial\theta} \eqno(4.16)$$

\noindent for any arbitrary fermionic gauge fixing term. Once again,
being the $\theta$ camponent of a superfield, this is manifestly invariant
under the extended BRST transformations of Eq. (2.11).

The complete Lagrangian can now be written as (see ref. [7] for a
discussion on the structure of the classical Lagrangian ${\cal L}_o$)

$$\eqalign{ \hat {\cal L} &= {\cal L}_o(\phi_\mu - \tilde\phi_\mu ) +
\tilde {\cal L}_{gf} + {\cal L}_{gf}\cr
\noalign{\vskip 4pt}%
&= {\cal L}_o (A_\mu - \tilde A_\mu ) +  {\partial\over \partial\theta}
\ {\rm Tr}\  \Big( \tilde\phi_\mu^*\tilde\phi^\mu
+ \tilde{\overline \eta}^* \tilde\eta -
+ \tilde{\overline \eta} \tilde\eta^*
-\tilde f^* \tilde  f\Big) + {\partial\Psi \over \partial\theta} \cr}
\eqno(4.17)$$

\noindent which is manifaestly invariant under the extended BRST symmetry
and upon elimination of the auxiliary fields and ghosts associated
with the shift symmetry, leads to the BV action.

\vskip 1cm
\noindent{\bf V. \underbar{Extended BRST and anti-BRST Invariant
Superspace Formulation  }}

\medskip

Let us recall briefly some facts about the superspace formulation of
a gauge theory with BRST and anti-BRST invariance [7]. If we define a
superspace with coordinates $(x^\mu, \theta,
\overline\theta)$, then in such a space, we can define a super
connection 1-form of the kind

$$ \omega = \phi_\mu (x,\theta,\overline\theta) dx^\mu +
\eta (x,\theta,\overline\theta) d\theta +
\overline\eta (x,\theta,\overline\theta) d\overline\theta \eqno(5.1)$$

\noindent where $ \phi_\mu,\ \eta$ and $\overline\eta$ are matrices
belonging to the adjoint representation of the gauge group SU(n).
The field strength, in this case, is given by

$$ F = d\omega + {1\over 2}\ [\omega , \omega ] \eqno(5.2)$$

\noindent and will have components not only along the $\mu,\nu$
directions, but also along all possible $\theta ,\  \overline\theta$
directions. Requiring the field strength to vanish along all extra
directions determines the superfields uniquely to be

$$\eqalign{  \phi_\mu (x,\theta,\overline\theta) &= A_\mu(x) +
\theta D_\mu c + \overline\theta D_\mu \overline c
+ \theta\overline\theta (D_\mu F - [D_\mu c , \overline c ]_+ )\cr
\noalign{\vskip 4pt}%
\eta (x,\theta,\overline\theta) &= c(x) - {1\over 2}\ \theta [c,c]_+
+ \overline\theta (F- [c,\overline c]_+ ) + \theta\overline\theta
(-[F,c] + [\overline c, c^2])\cr
\noalign{\vskip 4pt}%
\overline\eta (x,\theta,\overline\theta) &= \overline c(x) - \theta F
-{1\over 2}\ \overline\theta [\overline c,\overline c]_+
- \theta\overline\theta [F,\overline c]  \cr}\eqno(5.3)$$

It is now straight forward to compare with Eqs. (2.4) and (2.7) and
note that the BRST and anti-BRST transformations merely correspond,
in this formulation, to translations in the $\theta$ and
$\overline\theta$ coordinates respectively. The complete Lagrangian
of a gauge theory, with BRST and anti-BRST invariant gauge fixing,
can be written in this superspace where the invariances are manifest.

Let us next consider a Lagrangian density

$$\tilde {\cal L} = {\cal L} (  \phi_\mu - \tilde \phi_\mu,
\eta - \tilde \eta ,  \overline\eta -  \tilde {\overline\eta} )
\eqno(5.4)$$

\noindent which reduces to the original Lagrangian density when all the
tilde superfields vanish. Let us now define

$$\eqalign{  \Phi_\mu (x,\theta,\overline\theta) &=
\phi_\mu (x,\theta,\overline\theta) -
\tilde\phi_\mu (x,\theta,\overline\theta)\cr
\Lambda (x,\theta,\overline\theta) &=
\eta (x,\theta,\overline\theta) -
\tilde\eta (x,\theta,\overline\theta)\cr
\overline\Lambda (x,\theta,\overline\theta) &=
\overline\eta (x,\theta,\overline\theta) -
\tilde {\overline\eta} (x,\theta,\overline\theta)
  \cr}\eqno(5.5)$$

\noindent These new superfields clearly coincide with the original
superfields when the tilde superfields vanish.

If we now define the connection 1-form in this superspace as

$$\Omega = \Phi_\mu dx^\mu + \Lambda d\theta +
\overline\Lambda d\overline\theta \eqno(5.6)$$

\noindent and require the components of the field strength along all
the extra directions to vanish, then, once again we will determine

$$\eqalign{  \Phi_\mu (x,\theta,\overline\theta) &=
( A_\mu - \tilde A_\mu )  +
\theta  D_\mu^{(A-\tilde A)}( c -\tilde c)
 + \overline\theta D_\mu^{(A-\tilde A)}
( \overline c - \tilde{\overline c})\cr
\noalign{\vskip 4pt}%
&\qquad + \theta\overline\theta (D_\mu^{(A-\tilde A)}( F -\tilde F)
- [D_\mu^{(A-\tilde A)}( c -\tilde c) ,
( \overline c- \tilde{\overline c)} ]_+ )\cr
\noalign{\vskip 4pt}%
\Lambda (x,\theta,\overline\theta) &= (c - \tilde c)
 - {1\over 2}\ \theta [c- \tilde c,c- \tilde c]_+
+ \overline\theta ((F- \tilde F)
- [c -\tilde c,\overline c - \tilde{\overline c}]_+ ) \cr
\noalign{\vskip 4pt}%
&\qquad + \theta\overline\theta
\Big( -[F -\tilde F, c - \tilde c] + [\overline c - \tilde{\overline c},
(c-\tilde c)^2] \Big) \cr
\noalign{\vskip 4pt}%
\overline\Lambda (x,\theta,\overline\theta) &=
(\overline c -\tilde{\overline c})
 - \theta (F -\tilde F )
-{1\over 2}\ \overline\theta [\overline c-\tilde{\overline c} ,
 \overline c-\tilde{\overline c} ]_+\cr
\noalign{\vskip 4pt}%
&\qquad - \theta\overline\theta [F -\tilde F,\overline c
-\tilde{\overline c}]  \cr}\eqno(5.7)$$

The individual superfields  $\phi_\mu ,\ \eta ,\  \overline\eta$ and
$\tilde\phi_\mu, \  \tilde\eta ,\  \tilde {\overline\eta} $, however,
 will contain
arbitrary functions. This is the reflection of the arbitrariness in
the extended BRST and anti-BRST transformations that we discussed earlier.
Consistent with the discussion in sections II and III, we can choose
the individual superfields as follows.

$$\eqalign{  \phi_\mu (x,\theta,\overline\theta) &=
 A_\mu  + \theta \psi_\mu +
 + \overline\theta \big( A^*_\mu +  D_\mu^{(A-\tilde A)}
( \overline c - \tilde{\overline c}) \big) \cr
\noalign{\vskip 4pt}%
&\qquad + \theta\overline\theta (B_\mu + D_\mu^{(A-\tilde A)}( F -\tilde F)
- [D_\mu^{(A-\tilde A)}( c -\tilde c) ,
( \overline c- \tilde{\overline c)} ]_+ )\cr
\noalign{\vskip 4pt}%
\tilde\phi_\mu (x,\theta,\overline\theta) &=
 \tilde A_\mu  +
\theta \big( \psi_\mu -  D_\mu^{(A-\tilde A)}( c -\tilde c)\big)
 + \overline\theta A^*_\mu
+ \theta\overline\theta B_\mu\cr
\noalign{\vskip 4pt}%
\eta (x,\theta,\overline\theta) &= c +
\theta \epsilon
+ \overline\theta (c^* + (F- \tilde F)
- [c -\tilde c,\overline c - \tilde{\overline c}]_+ ) \cr
\noalign{\vskip 4pt}%
&\qquad + \theta\overline\theta
(B -[F - \tilde F,c - \tilde c] + [\overline c -
\tilde{\overline c}
, (c - \tilde c)^2])\cr
\noalign{\vskip 4pt}%
\tilde\eta (x,\theta,\overline\theta) &=  \tilde c
+\theta \bigg( \epsilon + {1\over 2}\ [c- \tilde c,c- \tilde c]_+\bigg)
+ \overline\theta c^*
+ \theta\overline\theta B\cr
\noalign{\vskip 4pt}%
\overline\eta (x,\theta,\overline\theta) &=
\overline c
+ \theta \overline\epsilon
+ \overline\theta \bigg( \overline c^* - {1\over 2}\
 [\overline c-\tilde{\overline c} ,
 \overline c-\tilde{\overline c} ]_+ \bigg)\cr
\noalign{\vskip 4pt}%
&\qquad + \theta\overline\theta (\overline B -  [F -\tilde F,\overline c
-\tilde{\overline c}] ) \cr
\noalign{\vskip 4pt}%
\tilde{\overline\eta} (x,\theta,\overline\theta) &=
\tilde{\overline c}
+ \theta (\overline\epsilon + (F - \tilde F ) )
+ \overline\theta \overline c^*
+ \theta\overline\theta \;\overline B  \cr }\eqno(5.8)$$

\noindent Here $\psi_\mu ,\  \epsilon ,\ \overline\epsilon,\
 A^*_\mu ,\  c^*,\
\overline c^* ,\  B_\mu,\  B$ and $\overline B$ are arbitrary functions and
can be identified with the ghosts, antighosts, and auxiliary fields
introduced in sec. II in connection with the shift symmetry.
In fact, it is straight forward to see that translations in the
$\theta$ and $\overline\theta$ coordinates generate respectively the
extended BRST and anti-BRST transformations of Eqs. (2.11) and (3.1) with

$$F^* =0 = B_F \eqno(5.9)$$

\noindent Furthermore, Eq. (5.8) determines the variations only in the
combination $(F-\tilde F ) $ and consequently, the ghost field,
$\epsilon_F$, does not occur in the superfields. We note here that
in a non gauge theory, one can introduce an additional superfield

$$ \tilde f (x,\theta,\overline\theta) = \tilde F +\theta
\epsilon_F + \overline\theta F^* + \theta\overline\theta B_F
\eqno(5.10)$$

\noindent to generate the additional transformations in Eqs. (2.11)
and (3.1). In a gauge theory, however, the auxiliary fields occur in
the antighost multiplet and, therefore, this will be artificial and
would destroy the geometrical formulation of the gauge theory.  We
also note here that the additional transformations (or
generalizations of them) can probably be generated by expanding the
tilde field structures [5]. But we feel that the minimal structure in
Eq. (5.8) is quite appealing and as we will show shortly the absence of
the variables in Eq. (5.9) does not pose any particular difficulty  in
the construction of the theory.

To proceed, let us note from the structure of the superfields in
Eq. (5.8) that

$$\eqalign{&-{1\over 2}\  {\partial\over \partial\overline\theta}\
{\partial\over \partial\theta}\ {\rm  Tr}\  \tilde\phi_\mu\tilde\phi^\mu =
\ {\rm Tr}\  \big( -B_\mu\tilde A^\mu - A^*_\mu (\psi^\mu -
D^{\mu (A-\tilde A)} (c-\tilde c)) \big)\cr
&{\partial\over \partial\overline\theta}\
{\partial\over \partial\theta}\ {\rm  Tr}\  \tilde\eta\tilde{\overline \eta} =
\ {\rm Tr}\  \bigg( -\overline B\tilde c + \overline c^* \bigg(
\epsilon + {1\over 2}\
[c-\tilde c, c-\tilde c]_+ \bigg)\cr
&\qquad\qquad\qquad\qquad\qquad\qquad + B\tilde{\overline c} -
c^* \big( \overline \epsilon + (F-\tilde F)\big) \bigg)
 \cr}\eqno(5.11)$$

\noindent Consequently, we can write

$$\eqalign{ \tilde {\cal L}'_{gf} &= \ {\rm Tr}\  \bigg[ -B_\mu\tilde A^\mu
- A_\mu^* (\psi^\mu -
D^{\mu (A-\tilde A)} (c-\tilde c) ) - \overline B \tilde
c + \overline c^* \bigg( \epsilon + {1\over 2}\  [c-\tilde c,c - \tilde c]_+
\bigg)
\cr
&\;\;\;\;\;\;\;\;
+ B \tilde {\overline c} - c^* (\overline \epsilon + (F- \tilde F  ))
\bigg] =  {\partial\over \partial\overline\theta}\
{\partial\over \partial\theta}\  {\rm Tr}\  \bigg( -{1\over 2}\
\tilde\phi_\mu\tilde\phi^\mu
+ \tilde\eta\tilde{\overline \eta}\bigg)\cr }\eqno(5.12)$$

\noindent Being the $\theta\overline\theta$ component of a superfield, this
gauge fixing Lagrangian is manifestly invariant under extended BRST
and anti-BRST transformations. Note that the gauge fixing Lagrangian
density in (5.12) differs from that in (2.12) in the $F$ dependent terms.
We will come back to this point shortly.

To write the gauge fixing Lagrangian for the original symmetry, we
note that we can choose a fermionic superfield whose first component
corresponds to the arbitrary fermionic gauge fixing function of Eq. (2.4).
Then, we have

$$  \Psi (x,\theta,\overline\theta) = \psi + \theta\delta\psi
+ \overline \theta \; \overline\delta\psi
+ \theta\overline\theta \delta\overline\delta \psi
\eqno(5.13) $$

\noindent In general, all four components of the superfield will be
non trivial implying that if we choose as in Eq. (2.4)

$${\cal L}_{gf} = \delta \psi  $$

\noindent then it will not be invariant under extended anti-BRST
transformations. (This follows from the fact that the $\theta
\overline\theta$ component of the superfield in Eq. (5.13) is non
vanishing in general.) However, we note that if the gauge fixing
Lagrangian is both BRST and anti-BRST invariant, then the
 $\theta \overline\theta$ component of $  \Psi (x,\theta,
\overline\theta)$ would vanish on-shell because when we use the equations
of motion, the tilde fields vanish and the theory reduces to the
original theory, where, by assumption the gauge fixing Lagrangian is
both BRST and anti-BRST invariant. This can, of course, be explicitly
checked for specific gauge choices. Thus, for example, for the
covariant gauge fixing of Eq. (2.6), we can write

$$\eqalign{  \Psi (x,\theta,\overline\theta) = &- {\rm Tr}\
\partial_\mu\overline\eta \phi^\mu = \ {\rm Tr}\ \Big( -\partial_\mu
\overline c A^\mu -  \theta (\partial_\mu \overline \epsilon
A^\mu - \partial_\mu\overline c \psi^\mu ) \cr
&- \overline \theta \big( ( \partial_\mu\overline c^* -
[\partial_\mu(\overline c - \tilde {\overline c} ),
\overline c - \tilde {\overline c} ]_+ )A_\mu
- \partial_\mu\overline c (A^{\mu *} + D^{\mu(A-\tilde A)}
(\overline c - \tilde {\overline c}))\big)\cr
&- \theta\overline\theta \big( (\partial_\mu \overline B -
[\partial_\mu (F-\tilde F), \overline c - \tilde {\overline c}] -
[F-\tilde F,\partial_\mu (\overline c - \tilde {\overline c}) ]
\big) A^\mu\cr
 &+ \partial_\mu\overline c \big( B^\mu + D^{\mu(A-\tilde A)}(F-\tilde
F) - [  D^{\mu(A-\tilde A)} (c-\tilde c),\overline c -
\tilde {\overline c}]_+\big)\cr
&-(\partial_\mu\overline c^* - [\partial_\mu
(\overline c - \tilde {\overline c}),\overline c -
\tilde {\overline c}]_+ )\psi^\mu +
\partial_\mu\overline\epsilon ( A^{\mu *} +
 D^{\mu(A-\tilde A)}
(\overline c - \tilde {\overline c}) )\Big)}
\eqno(5.14) $$

If we choose as the gauge fixing Lagrangian the $\theta$ component of
this Lagrangian, namely (neglecting total divergences)

$${\cal L}_{gf} =  \partial_\mu\overline c \psi^\mu -
\partial_\mu \overline \epsilon A^\mu \equiv
 \partial_\mu\overline c \psi^\mu +
\partial_\mu  A^\mu \overline \epsilon\eqno(5.15)$$

\noindent it is tedious but straight forward to check that the
$\theta\overline\theta$ component in Eq. (5.14) vanishes when we use
the equations of motion.  Therefore, it is clear that for an
arbitrary fermionic gauge fixing function that leads to a BRST and
anti-BRST gauge fixing Lagrangian, we can choose

$${\cal L}_{gf} = {\partial\over \partial\theta}\
\big( \delta(\overline\theta) \Psi (x,\theta,\overline\theta)\big)
\eqno(5.16)$$

\noindent This would, of course, be manifestly invariant under
extended BRST transformations, but it will also be invariant under
extended anti-BRST transformations on-shell. One can presumably add
necessary auxiliary fields to make the gauge fixing Lagrangian
invariant under extended anti-BRST transformations without the use of
the equations of motion. However, this would take us out of the
minimal geometric approach, which is the spirit of our paper.

The complete Lagrangian, which is invariant under extended BRST
transformations and is also invariant under extended anti-BRST
transformations on-shell can, therefore, be written as

$$\eqalign{\hat{\cal L} &=  {\cal L}_o(\phi_\mu - \tilde\phi_\mu ) +
\tilde {\cal L}_{gf}^\prime + {\cal L}_{gf}\cr
\noalign{\vskip 4pt}%
&= {\cal L}_o (A_\mu - \tilde A_\mu ) +
 {\partial\over \partial\overline\theta}\
{\partial\over \partial\theta}\ {\rm  Tr}\  \bigg( -{1\over 2}\
\tilde\phi_\mu\tilde\phi^\mu
+ \tilde\eta\tilde{\overline \eta}\bigg)
+  {\partial\over \partial\theta} \
\big( \delta(\overline\theta) \Psi (x,\theta,\overline\theta)\big)\cr
\noalign{\vskip 4pt}%
&= \ {\rm Tr}\ \bigg( -{1\over 4}\ F_{\mu\nu}(A-\tilde A)F^{\mu\nu}(A-\tilde A)
 -B_\mu\tilde A^\mu  - \overline B \tilde c + B \tilde {\overline c}\cr
&\qquad - \bigg( A_\mu^* +{\delta \psi\over \delta A^\mu} \bigg) \psi^\mu
+ \bigg( \overline c^* +{\delta\psi\over \delta c}\bigg) \epsilon
 - \bigg( c^* - {\delta \psi\over \delta \overline c}\bigg) \overline \epsilon
\cr
&\qquad + A_\mu^* D^{\mu (A-\tilde A)} (c-\tilde c)\cr
&\qquad + {1\over 2}\  \overline c^* [c-\tilde c,c - \tilde c]_+
 - c^*  (F- \tilde F  )\bigg) \cr }\eqno(5.17)$$

\noindent We note that integrating out the auxiliary fields $B_\mu ,\
\overline B$ and $B$ will set the tilde fields $\tilde A^\mu ,\  \tilde c$ and
$\tilde{\overline c}$ to zero. Furthermore, integrating out the gost
fields for the shift symmetry, $\psi^\mu,\  \epsilon$
and $\overline\epsilon$ will determine the antifields
$A^*_\mu ,\   c^* $ and $\overline c^* $ which, when substituted into
the Lagrangian density will yield the BV action except for the
$\tilde F$ term. But we note that $F$ and $\tilde F$ are auxiliary fields
and, therefore, one can  trivially redefine

$$ F - \tilde F \rightarrow F \eqno(5.18)$$

\noindent The orthogonal combination $(F+\tilde F)$ can be integrated
out of the functional integral leading to an infinite constant which
can be absorbed into the normalization of the path integral. With
these redefinition, then, the BV action is obtained.

\vskip 1cm

\noindent{\bf VI. \underbar{Conclusion }}

\medskip

The BV formalism provides a powerful quantization method within the
Lagrangian formulation. Here one extends the configuration space by introducing
anti-fields corresponding to all the original fields present in the
theory. (These are ultimately identified with the functional
derivative of a gauge fixing fermion with respect to the
corresponding fields.) In this space, one defines an antibracket
which generates the BRST transformations with the classical BV action
as the generator. The complete quantum action is obtained as a proper
solution of the master equation in a power series in $\hbar$ which
coincides with the classical BV action (up to renormalization) when
there are no anomalies present. In this paper we have considered such
a theory$^{[10]}$ and have shown that if the gauge fixing fermion leads to a
BRST and anti-BRST invariant Lagrangian, then the BV action
formulated with an extended BRST symmetry is also invariant under an extended
anti-BRST symmetry. We have tried to give a geometrical meaning to these
extended transformations as corresponding to translations in a superspace.
For a general gauge fixing fermion, we have shown that the BV action
can be written in a manifestly extended BRST invariant manner in a
superspace with one Grassmann coordinate. On the other hand, if the
gauge fixing fermion leads to a BRST and anti-BRST invariant
Lagrangian, then we have shown that the BV action can be written in a
maniifestly extended BRST and anti-BRST invariant manner (at least on
shell) in a superspace with a pair of Grassmann coordinates.
This formalism will readily generalize to other non anomalous
theories.  This geometric formulation is manifestly BRST invariant
and it will be interesting to see how anomalous gauge theories would fit
into this formulation. This is presently under study.

\vskip 1cm

\noindent {\bf \underbar{Acknowledgements}}

\medskip

 This work was supported in part by
U.S.Department of Energy Grant No. DE-FG-02-91ER40685 and by
CNPq - Brazil.
\vfill\eject

\noindent {\bf \underbar{References}}

\medskip

\item{1.} I. A. Batalin and G. A. Vilkovisky, Phys. Lett. {\bf B102}
(1981) 27; Phys. Rev. {\bf D28} (1983) 2567.
\item{2.} For a review see M. Henneaux and C. Teitelboim, Quantization
of Gauge Systems, Princeton University Press 1992, Princeton, New
Jersey and F. De Jonghe, \lq\lq The Batalin-Vilkovisky Lagrangian Quantization
Scheme with Applications to the Study of Anomalies in Gauge Theories",
Ph.D. thesis K. U. Leuven, HEP-TH 9403143.
\item{3.} C. Becchi, A. Rouet and R. Stora, Comm. Math. Phys.
{\bf 42} (1975) 127;
I. V. Tyutin, Preprint Lebedev Inst. No. 39 (1975).
\item{4.} G. Curci and R. Ferrari, Phys. Lett. {\bf B63} (1976) 51.
\item{5.} For alternate discussions of the anti-BRST symmetry in
the BV context see:
\item{  }  I. A. Batalin, P. M. Lavrov, I. V. Tyutin, J. Math. Phys.
{\bf 31} (1990) 1487; {\it ibid} {\bf 32} (1991) 532, {\bf 32} (1991) 2513;
 J. Gomis and J. Roca, Nucl. Phys. {\bf B343} (1990) 152 and also the last
 reference in [2].
\item{6.} S. Ferrara, O. Piguet and M. Schweda, Nucl. Phys.
{\bf B119}
(1977) 493;  K. Fujikawa, Progr. Theor. Phys. {\bf 59} (1977) 2045.
\item{7.} L. Bonora and M. Tonin, Phys. Lett. {\bf B98} (1981) 48.
\item{8.} J. Alfaro and P. H. Damgaard, Phys. Lett. {\bf B222}
 (1989) 425;
J. Alfaro, P. H. Damgaard , J. I. Latorre and D. Montano, Phys.
 Lett. {\bf B233}
(1989) 153; J. Alfaro and P. H. Damgaard, Nucl. Phys. {\bf B404}
 (1993) 751.
\item{9.} A. Das and M. A. Namazie, Phys. Lett. {\bf 99B}
(1981) 463; A. Das,
Phys. Rev. {\bf D26} (1982) 2774; L. Baulieu and J. Thierry-Mieg, Nucl. Phys.
{\bf B212} (1983) 255.
\item{10.} We have not been concerned with the question of regularization
here, since we have considered a non anomalous gauge theory. For
the case of anomalous gauge theories in the BV context, the regularization
plays an important role. See, for
example, W. Troost, P. van Nieuwenhuizen and A. Van Proeyen, Nucl.
Phys. {\bf B333} (1990) 727.

\bye